# Nonlinear Metasurfaces: A Paradigm Shift in Nonlinear Optics

Alexander Krasnok, Mykhailo Tymchenko, and Andrea Alù

Department of Electrical and Computer Engineering, The University of Texas at Austin, USA

## Abstract

Frequency conversion processes, such as second- and third-harmonic generation, are one of the most common effects in nonlinear optics which offer many opportunities for photonics, chemistry, material science, characterization, and biosensing. Given the inherently weak nonlinear response of natural materials, one typically relies on optically large samples and complex phase-matching techniques to achieve nonlinear effects. A direct translation of these approaches to small dimensions, however, is extremely challenging, because nonlinear effects locally comparable to the linear response cannot be induced without reaching extremely high light intensities leading to the material breakdown. For this reason, the quest to synthesize novel materials with enhanced optical nonlinearities at moderate input intensities is very active nowadays. In the last decade, several approaches to engineering the nonlinear properties of artificial materials, metamaterials, and metasurfaces have been introduced. Here, we review the current state of the art in the field of small-scale nonlinear optics, with special emphasis on high-harmonic generation from ultrathin metasurfaces, including those based on plasmonic and high-index dielectric resonators, as well as semiconductor-loaded plasmonic metasurfaces. We discuss the role of specific electromagnetic field configurations for efficient harmonic generation, including magnetic dipole, Fano resonance, and anapole ones. We also discuss the most recent advances in controlling the phase front profiles of generated nonlinear waves enabled by nonlinear metasurfaces consisting of nanoresonators with judiciously tailored shapes. Finally, we compare viable approaches to enhance nonlinearities without phase matching constraints in ultrathin metasurfaces, and offer a perspective and outlook on the future development of this exciting field.

**Keywords**

Nonlinear optics, nonlinear metasurfaces, plasmonic nanostructures, dielectric nanostructures, multi-quantum-wells

## I. Introduction and Background

Photon-photon interactions are intrinsically weak, and they can only be realized in the presence of matter at high light intensities [1]. The study of these interactions is the essence of nonlinear optics which is one of the bedrocks of modern photonic science and technology. When light is impinging upon a material, the collective displacement of electrons generates a scattered field oscillating synchronously with the impinging light and interfering with the original field. At low amplitudes, the scattered field is linearly proportional to the impinging one, and the interference between them is the origin of linear optical responses and weak light-matter interactions, such as refraction and reflection. However, if the amplitude of impinging light is sufficiently large, electronic bindings in the material become increasingly asymmetric, and bound electron orbits become distorted [1]. It means that the optical response of matter is now intensity-dependent, and the effects of nonlinear optics come to the scene. We may estimate light intensities that are needed to achieve the nonlinear optical response. Neglecting the saturation effects, they should be comparable with intra-atomic fields, i.e. on the order of $e/r_o^2 \approx 10^9$ V/cm, where $e$ is the electron charge, and $r_o \approx 10^{-8}$ cm is the characteristic radius of electronic orbits. In the general case, the orbit distortion and the resulting bulk material response can be described by a polarization vector **P** having both linear and higher order (nonlinear) terms,

$$P_i = \varepsilon_0 \sum_j \chi_{ij}^{(1)} E_j + \varepsilon_0 \sum_{jk} \chi_{ijk}^{(2)} E_j E_k + \varepsilon_0 \sum_{jkl} \chi_{ijkl}^{(3)} E_j E_k E_l + ... , \tag{1}$$

where $\mathbf{E}$ is a local field, $\varepsilon_0$ is a vacuum permittivity, $\chi^{(1)}$ is a linear susceptibility of the medium, and $\chi^{(2)}$, $\chi^{(3)}$ are the second- and third-order nonlinear susceptibility tensors, respectively. With an increase of applied field intensity, the nonlinear terms, $\chi^{(2)} E^2$ and $\chi^{(3)} E^3$, can become non-negligible, and the material response becomes nonlinear. Compared to linear optics which focuses exclusively on the first term in Eq. (1), nonlinear optics offers a rich and diverse set of nontrivial electromagnetic phenomena. For a monochromatic incident field $\mathbf{E} = \mathbf{E}^\omega \cos \omega t$, the polarization $\mathbf{P}$ has nonlinear terms oscillating at various frequencies, among which are $\omega$ (Kerr effect), $2\omega$ (second-harmonic generation, SHG), $3\omega$ (third-harmonic generation, THG), etc., as well as constant terms responsible for an intensity-dependent refractive index stemming from optical rectification. Similarly, a double-frequency impinging field $\mathbf{E} = \mathbf{E}^{\omega_1} \cos \omega_1 t + \mathbf{E}^{\omega_2} \cos \omega_2 t$, when substituted into (1), leads to nonlinear terms that oscillate at frequencies $\omega_1 \pm \omega_2$ (sum- and difference-frequency generation, SFG and DFG, respectively). This list is by no means complete, and many other interesting phenomena can arise in the nonlinear regime [1].

It is important to mention that, unlike linear effects intensities of which are always proportional to the incident field, higher-order nonlinear phenomena are also subject to symmetry constraints. For instance, the second term in Eq. (1) becomes zero is the structure is centrosymmetric [1]. This results comes from the fact that if the field $-\mathbf{E}$ is applied instead of $\mathbf{E}$ to a centrosymmetric crystal or structure, $\mathbf{P}$ must change its sign due to the spatial inversion, that yields: $-\mathbf{P}^{(2)} = \varepsilon_0 \chi^{(2)} (-\mathbf{E})^2$. After comparison with Eq. (1), we obtain $-\mathbf{P}^{(2)} = \mathbf{P}^{(2)}$, so that $\chi^{(2)}$ must be zero. This result is of fundamental importance since it explicitly forbids bulk crystals with centrosymmetric lattices to exhibit second-order nonlinear responses regardless of the intensity of the applied field. The same symmetry constraints apply to artificially designed structures aimed, for instance, at SHG, as discussed below. Third-order nonlinear effects, such as the Kerr effect, THG, and the four-wave mixing (FWM), are not bound by these constraints.

Frequency conversion in various materials has been a subject of extensive research for many years. Very recent applications includes nanostructured materials characterization [2], coherent ultraviolet light generation [3], [4], [5], supercontinuum white light generation [6], spectroscopy [7], bioimaging and sensing [8], nanomedicine [9], quantum optics (generation of entangled photon pairs) [10], [11], and broadening of the spectral range accessible with existing lasers. Building upon significant achievements in nonlinear optics based on conventional bulk materials, the modern research trend has shifted towards miniaturization of nonlinear optical components placed in a more compact setups. A direct translation of conventional nonlinear optics approaches to smaller dimensions, however, turned out to be extremely challenging. The reason lies primarily in the fact that intrinsic nonlinearities in natural materials are extremely weak ($\chi^{(2)} \approx 10^{-12}$ m/V, and higher-order terms are even smaller). Therefore nonlinear responses locally comparable to the linear response of optical materials cannot be induced without reaching very high intensities causing the material breakdown. This implies that large bulk volumes are necessary to allow significant nonlinear responses to build up. The use of bulk materials, often hundreds of wavelengths thick and exhibiting inherently dispersive refraction indexes, requires extremely precise and often cumbersome techniques to compensate for the momentum mismatch between pump and generated signals [1]. The complexity of these setups commonly involving dozens of bulky optical elements makes engineering of nonlinear optical devices cumbersome and inconvenient.

The issue of low nonlinear conversion efficiency in small volumes is being partially addressed by synthesizing new materials with specifically tailored large nonlinearities. Current approaches to engineering the nonlinear properties of materials rely mainly on chemistry approaches [12], and on the use of artificial materials and metamaterials [13]–[21].

The chemical approach allows researchers to create novel materials with unusual properties, including enhanced nonlinear response. For example, *metal-organic framework* compounds, i.e., structures consisting of metal ions or clusters coordinated to organic ligands, in the last few years have been attracting significant attention due to their enhanced second-order nonlinear response (see Ref. [22] and references therein). Other examples of chemically synthesized materials that draw a growing attention are organic-inorganic [23]–[25] and inorganic [26],[27] *perovskites*. These materials have also demonstrated efficient nonlinear harmonic generation [26]–[28]. However, despite this significant progress in material science, such bulk nonlinear media still require phase-matching, hindering their use in compact devices.

Another emerging class of nonlinear media is 2D materials, for instance single- or multi-layer graphene [28]–[33], $MoS_2$ [34]–[37], $MoSe_2$ [38], [39], boron nitride [37], and black phosphorus [40]. It turns out that even such vanishingly-thin structures can have very large nonlinear coefficients. The interest to studying optical nonlinearities in 2D materials is fueled by the fact that if their properties turn out to be superior, they can replace conventional nonlinear media in embedded photonic systems.

The artificial material approach has also led to a new class of artificial semiconductor media – *multi-quantum-wells* (MQWs), – exhibiting one of the largest bulk nonlinearities among all condensed-matter systems [41]–[43]. Typically, MQWs are composed of stacks of different III-V semiconductor layers serving as a set of potential wells and barriers for charge carriers moving in the transverse direction. Electrons and holes confined in these wells occupy discrete energy levels which can be precisely engineered by controlling the layer thicknesses (up to one atomic layer) using modern semiconductor growth technology. Band engineering in semiconductors has become an extremely fruitful and mature field of research, leading to numerous advances in a variety of research fields. Particularly in the context of nonlinear optics, MQWs have shown the possibility to engineer very large second- and third-order susceptibilities, 4–5 orders of magnitude larger than those of any natural material [43],[44]. MQWs, therefore, enable the use of smaller, potentially subwavelength volumes, that significantly relaxes phase-matching requirements. However, for optical applications MQWs possess a significant drawback: their nonlinear response can be achieved only for light polarized normally to the layers. Thus, some sort of clever structure engineering is required in order to efficiently couple light in and out of the layers to engage the MQW nonlinearity.

In parallel with this progress at the material engineering level, new strategies simultaneously addressing efficiency and phase-matching have been devised and pursued over the last decade. A promising and rapidly advancing approach to efficient nonlinear generation is following the metamaterial paradigm, i.e. constructing artificial materials with exotic electromagnetic responses. In metamaterials, collections of metallic or high-index dielectric nanoparticles, holes, and cavities enable dramatically enhanced light-matter interactions at subwavelength scales by controlling and tailoring local polarizations, phases, and amplitudes of linear and nonlinear local fields [45]–[52]. A primary goal of metamaterials and their 2D counterparts, metasurfaces, – obtaining specific optical properties on demand – is achieved by judicious engineering the shape and arrangement of these sub-diffractive meta-atoms [19],[50],[53]–[64]. For instance, one approach to enhancing the nonlinear response is based on the use of epsilon-near-zero (ENZ) materials and metamaterials [65]–[69]. ENZ media are materials with the real part of permittivity near zero at the frequency of interest. Wave propagation in these materials is very peculiar, as they exhibit very large wavelengths and, therefore, a nearly static spatial field distribution with no phase advance.

Although ENZ condition can occur naturally, as in the case of SiC at mid-infrared frequencies or the ionosphere in the radio-frequency range, it is usually achieved in waveguide-like structures operated near the cutoff frequency. ENZ materials enable interesting linear phenomena including tunneling and squeezing of energy through arbitrarily distorted narrow channels [70], highly directive emission from light sources [71], tailoring of the radiation phase pattern [72], trapping of light [73], and strong Purcell enhancements [74]. Moreover, ENZ materials are becoming a platform for obtaining non-classical, non-reciprocal, and non-local responses of matter [67], [75], [76]. ENZ structures loaded with nonlinear media are particularly well-suited for enabling coherent superposition of nonlinear effects over long distances, as they enable uniformly enhanced fields and a nearly homogeneous phase profile, that frees from phase-matching requirements and leads to enhanced nonlinear effects, in particular, Kerr nonlinearity [77],[79], SHG [66],[78], THG [77], and FWM [80].

The primary subject of this review are flat 2D *nonlinear metasurfaces*, which are much easier to realize than 3D geometries. In contrast to 3D structures, nonlinear metasurfaces can exhibit strong nonlinear optical responses in a much more compact footprint, that relaxes or completely overcomes phase-matching requirements. In the following, we take a close look at the state-of-the-art advances in currently booming area of flat nonlinear optics based on various 2D geometries which exhibit record-high nonlinear per-volume conversion efficiencies. We devote special attention to plasmonic (Section II), high-index all-dielectric (Section III), and combined metal-dielectric metasurfaces (Section IV), as is shown in Figure 1. Nonlinear metasurfaces can be aimed at various nonlinear processes that include SHG [81]–[95], THG [96]–[104], spontaneous parametric down-conversion [105]–[108], sum- and difference-frequency generation (SFG and DFG, respectively) [21], [109], [110], FWM [80], [111]–[116], nonlinear phase control and optical activity [117]–[124], nonlinear switching and routing [125]–[130], as is schematically sketched in

Figure 2. In this figure, the nonlinear signals are shown to be radiated in the forward direction ("transmission" regime). In many papers, however, the nonlinear signal is radiated in the reverse direction ("reflection" regime), which does not change the underlying physics. Given this very rich set of possible configurations, it is evident that flat and thin nonlinear metasurfaces constitute a novel, unified, and capable platform for all kinds of nonlinear processes. Section V is fully devoted to discussion of powerful linear and nonlinear field-tailoring capabilities provided by nonlinear metasurfaces. Due to strict phase-matching requirements, the subwavelength control of nonlinear fields in bulk nonlinear crystals used to be hardly achievable. On the contrary, building on the great success of their linear counterparts, nonlinear metasurfaces can sustain *nonlinear* polarization currents with deliberately controlled phase profile, further enriching their functionalities.

Finally, in the concluding Section VI, we summarize the main results on giant optical nonlinearities in metasurfaces, we compare the different approaches and their implementation, and we offer further perspectives on their future development. The main goal of this review is to show that nonlinear metasurfaces are causing a paradigm shift in nonlinear optics. They are poised to become a breakthrough in the quest to truly miniaturized, robust, and efficient nonlinear optical devices.

## II. Nonlinear Plasmonic Metasurfaces

Obtaining a strong nonlinear response from optically thin structures requires much stronger light-matter interactions than those that are naturally available in bulk nonlinear media. The use of surface plasmons existing at metal-dielectric interfaces is an efficient approach to enhance light-matter interactions and confine strong fields at the metallic boundary [51], [131]–[138]. The surface of virtually any material can support a nonlinear response, because electrons at the surface reside in a non-symmetric environment, and can therefore avoid the symmetry constraints, as was outlined in the Introduction section. In this case, the nonlinearities stem from the asymmetry of the potential confining the electrons at the material's surface [139]–[142]. Even purely metallic media with crystalline lattices which do not possess intrinsic bulk nonlinearity have been shown to emit nonlinear signals from the surface [143]. These effects can be dramatically enhanced if surface plasmons are supported by the interface, and if they are excited during the nonlinear process, e.g., by interface patterning [144]–[150] or a probing tip [151]–[153]. In contrast to bulk nonlinear media in which polarization and phase control is hard to achieve, the fact that surface plasmons at optical and mid-infrared frequencies are tightly bound to the metal interface, and that they follow its profile as they are guided, enables a large degree of control of the optical nonlinearity by engineering subwavelength plasmonic meta-atoms with precise polarization control and strong local field enhancement [154], [155].

Plasmonic metasurfaces open a plethora of interesting possibilities for nonlinear signal generation. The shape of these structures can be tailored to resonantly enhance light-matter interaction at specific frequencies, boosting the efficiency of nonlinear generation [47], [126], [149], [156]. Moreover, the deeply subwavelength thickness of nonlinear metasurfaces relaxes the necessity of considering cumbersome phase-matching conditions in the longitudinal direction. It is important to mention, however, that similarly to bulk nonlinear crystals, second-order nonlinear effects, such as SHG, in metasurfaces are subject to the same symmetry constraints, as discussed in the introduction. Thus, SHG in metasurfaces cannot be achieved in setups involving centrosymmetric modes' profiles both at $\omega$ and $2\omega$.

One way to achieve symmetry breaking required for SHG metasurfaces is by using plasmonic nanostructures with removed inversion symmetry, such as L- [84], [85], [157]–[160] and G-shaped antennas [2], [161]–[164], split-ring resonators [86], [165]–[169], asymmetric dimers [170], [171], dielectric-loaded plasmonic nanocups [82], and multi-resonant antennas [83], [91], [172], [173].

The nonlinear response in some of these structures, such as G-shapes and SRRs, has been at least partially attributed to magnetic resonances [20], [81], [163], [167], [174], [175]. However, metasurfaces of complementary geometries, such as metallic films with holes, for which the roles of electric and magnetic resonances are swapped, have been shown to produce similar SHG responses [165]. Detailed full-numerical studies of electric dipolar, quadrupolar and magnetic dipolar contributions to SHG processes in stand-alone spherical particles [176], [177] and their assemblies [173], as well as L-shaped nanoantennas [84], [176], have been reported in the recent literature. Yet, in more complex periodic structures, the roles of different resonances in nonlinear generation have not been unambiguously distinguished [126], [178]. For this reason, various design schemes and approaches are being pursued in this area.

Recently, it has been shown that dipolar, quadrupolar and higher-order contributions to nonlinear generation in nanostructures can be accurately taken into account by applying Lorentz reciprocity [179]. This approach revealed that, at least in the undepleted pump approximation, i.e., under the assumption that the overall nonlinear signal is sufficiently weak, the effective nonlinear response of the metasurface is proportional to the spatial overlap integral among fields excited in the structure by probing waves at pump and SH frequencies with specific polarization combinations, weighed by the local nonlinear susceptibility tensor and averaged over the unit-cell volume $V$ [169], [180]–[185]:

$$\chi^{(2)}_{\text{eff},lmn} = \frac{1}{V} \sum_{ijk} \int_V \chi^{(2)}_{ijk}(\mathbf{r}) \frac{E^{2\omega}_{i(l)}(\mathbf{r}) E^{\omega}_{j(m)}(\mathbf{r}) E^{\omega}_{k(n)}(\mathbf{r})}{E^{2\omega}_{\text{inc},l} E^{\omega}_{\text{inc},m} E^{\omega}_{\text{inc},n}} d^3\mathbf{r}. \qquad (2)$$

Here $E^{\omega}_{i(l)}$ is the *l*-polarized component of the field excited in a metasurface by a *l*-polarized incident probing wave, $E^{\omega}_{\text{inc},l}$. The effective nonlinear susceptibility of the metasurface $\chi^{(2)}_{\text{eff},lmn}$ relates the effective averaged second-order polarization of a unit cell and the *incident* field

$$P_{\text{eff},l}^{2\omega} = \varepsilon_0 \sum_{mn} \chi_{\text{eff},lmn}^{(2)} E_{\text{inc},m}^{\omega} E_{\text{inc},n}^{\omega} . \tag{3}$$

It is important to mention that Eqs. (2) and (3) are valid only in the quasi-static condition reached for unit-cells that are significantly smaller than the wavelength at $2\omega$. If this condition is not met, the volumetric averaging is not valid, and a more rigorous modelling technique must be employed to rigorously analyze SHG in a such metasurface. From Eq. (2), it is evident that, in order to maximize the second-harmonic response, the structure must provide strong field enhancement both at $\omega$ and $2\omega$, as shown in recent experimental studies [167]. This makes physical sense, since the structure needs to be able to sustain strong fields at the pump frequency, and at the same time be able to efficiently radiate the nonlinearly generated fields at the second harmonic frequency. Similar expressions can be derived for other nonlinear processes of choice. The averaging over the unit-cell volume in Eq. (2) automatically removes the dipolar contribution to SHG forbidden by the symmetry considerations outlined above. For metallic structures, the nonlinearity can arise only at the surface, since the fields rapidly decay as they enter the metal, and thus the overlap reduces to a surface overlap integral over the metal boundary or over the skin depth. If all fields are centrosymmetric, the overlap integral is identically zero because positive and negative portions cancel each other out. In turn, nanostructures with removed central symmetry can exhibit a spatially asymmetrical field overlap leading to the second-order nonlinear response.

The Eq. (2) gives us a set of requirements for a nanostructure so can exhibit the second-order nonlinear response: (i) a single cell resonator must lack inversion symmetry, (ii) the resonator must be designed to induce and enhance the local electric field at both fundamental frequency (FF) $\omega$ and the second-harmonic frequency (SH) $2\omega$. Several successful examples of nanoparticles that meet these requirements are shown in Figure 3(a-d). Interestingly, while asymmetry is definitely beneficial for SHG, breaking the centrosymmetry of the structure is neither required nor sufficient to achieve some form of SHG. For example, SHG has been observed in centrosymmetric structures, such as slightly rotated nanocrosses arranged in a closely packed array, see Figure 3(e) [186]. Being illuminated by a circularly polarized wave, this metasurface breaks the field overlap symmetry due to cell coupling and periodicity. Following the same idea, the overlap integral can also be made non-zero in symmetric structures such as arrays of metallic spheres, slits, and particle assemblies, see Figure 3(f), if the metasurface is excited and/or observed at oblique angles [4], [96], [172], [182], [187]–[191]. In such setups, the non-zero overlap emerges from spatial gradients of the field profile along the metasurface. Some experiments have reported significant SHG from centrosymmetric plasmonic nanocavities and apertures even at normal incidence and for linearly polarized fields [192]. In this case, the second-order response and symmetry breaking can be explained by the excitation of longitudinal field components in the waveguides near the cutoff. This explanation is supported by a strong correlation between SHG efficiency and the cutoff frequency of these apertures [126], [193].

For the sake of comparison, we note that for all setups described so far typical peak SHG power conversion efficiencies $\eta_{\text{SHG}} = P_{\text{SHG}}/P_{\text{FF}}$ ($P_{\text{FF}}$ is the pump power at the fundamental frequency (FF)), from purely metallic metasurfaces are in the order of $10^{-9}$ at $P_{\text{FF}} = 10-50$ μW (low enough to exclude any photodamage) [4], [83], [91], [165], [173], [175], [193]. After taking into account photon counting corrections, higher estimated values in the range of $10^{-6}$ have also been reported [167], [173]. In an ideal scenario, $P_{\text{SHG}}$ scales as $P_{\text{FF}}^2$, and thus potentially even larger $\eta_{\text{SHG}}$ may be achievable. However, substantial losses in metals lead to relatively low optical damage thresholds that fundamentally limit the SHG conversion efficiency in such structures. Similarly to SHG operation, plasmonic structures such as slits, nanogaps and holes in metallic films, have been shown to exhibit strong third-order nonlinear processes [194]–[196]. Excitation and focusing of surface plasmons at a metallic nanotip has been shown to lead to a dramatic increase in four-wave mixing efficiency [116], [197]–[199]. Finally, spiral plasmonic antennas has been used for supercontinuum white light generation [6].

To conclude this section, plasmonic nonlinear metasurfaces have recently become a major milestone on the route towards highly efficient nonlinear generation. Plasmonic nanostructures offer a robust platform for compact nonlinear devices producing a whole variety of nonlinear responses. Freed from phase-matching constraints, and featuring an unprecedented degree of control over nonlinear fields, plasmonic metasurfaces bring closer to practice the engagement of nonlinear processes in truly compact setups. However, due to relatively low thermal damage thresholds [200], [201], achieving substantially higher SHG conversion efficiencies in purely metallic structures is unlikely. In order to overcome this difficulty, the use of high-index all-dielectric nonlinear metasurfaces, discussed in the next section, is one of the promising approaches to achieve similar functionalities in higher-power setups.

### III. Nonlinear All-Dielectric Metasurfaces

In the previous section, we discussed how magnetic resonances in plasmonic structures can be employed to enable and enhance nonlinear processes, in addition to electric ones. In fact, under certain circumstances the efficiency of nonlinear generation from magnetic resonances can even dominate over the electric ones. However, plasmonic nanostructures have several disadvantages that limit their applicability for nonlinear nanophotonics, including high dissipative losses and inevitable thermal heating, leading to low optical damage thresholds. Thus, the use of all-dielectric metasurfaces supporting magnetic resonances, and able to withstand much higher pump field intensities, is a promising route to obtain higher nonlinear conversion efficiencies. Recently, it has been demonstrated that high-index dielectric nanoparticles can exhibit electric and magnetic dipole Mie-type resonances in the visible range [202]–[207]. Researchers have considered crystalline silicon (c-Si) with refractive index $n = 3.486...4.293$ and extinction coefficient $k = 0.001...0.045$ in the wavelength range $0.50...1.45\,\mu m$, and less often, GaAs ($n = 3.679...4.037$, $k = 0.089...0.376$) and Ge ($n = 4.460...5.811$, $k = 1.389...2.366$). Due to the relatively large values of refractive indexes, such nanoparticles support intriguing linear effects, such as directional scattering, absence of backward scattering, and power pattern steering. Moreover, it has been shown that, by tuning the electric and magnetic response of all-dielectric nanostructures, highly efficient and flexible light manipulation can be achieved at the nanoscale. The tailoring of the linear optical properties in these subwavelength systems is possible because the scattering cross-section strongly depends on the wavelength [208]–[211], the shape [212]–[214] and the optical properties of the substrates [215], [216].

High-index dielectric nanoparticles have recently become a promising platform for nonlinear nanophotonics, providing opportunities especially for ultrafast optical switching [128], [129], [217], [218] and high harmonic generation [5], [27], [94], [219], [220]. In this section, we present an overview of state-of-the-art advances in this area, discussing how all-dielectric nanostructures and metasurfaces may support even stronger nonlinear optical responses than their plasmonic counterparts, given their lower damage threshold that allows operation with high pump intensities [5], [221]. Moreover, in contrast to plasmonic nanoresonators, the electric field confined in dielectric nanoresonators is not limited to the surface. Thus, despite the fact that the electric field enhancement in dielectric nanoparticles is typically smaller than in plasmonic ones, we can expect an overall larger enhancement stemming from the volume resonance [219]. The fundamental role of mode volume has been highlighted in several types of nonlinear optical processes, including nanoparticle-enhanced Raman scattering [222].

Silicon (Si) has been considered as an excellent material platform for third-order nonlinear all-dielectric metasurfaces, given its large linear refractive index ($n \approx 3.7$) and intrinsic third-order susceptibility and nonlinear index ($\chi_{Si}^{(3)} \approx 2.79 \times 10^{-18} \, m^2/V^2$, $n_{2(Si)} \approx 2.7 \times 10^{-18} \, m^2/W$) in the near-infrared band [223], [224]. Germanium (Ge) is another excellent material, because of its high refractive index ($n_{Ge} \approx 4.460 - 5.811$, $k_{Ge} \approx 1.389 - 2.366$) in the visible range and large third order susceptibility ($\chi_{Ge}^{(3)} \approx 5.65 \times 10^{-19} \, m^2/V^2$) [97]. In contrast to third-order nonlinearity, silicon and germanium do not possess bulk-mediated second-order nonlinearity. In order to achieve strong second order nonlinearity, one can use GaAs, with $\chi_{GaAs}^{(2)} \approx 200 \, pm/V$ [225].

The first demonstration of strong nonlinear response in dielectric nanoparticles has been reported in Ref. [219], showing enhanced third-harmonic generation from silicon (Si) nanodisks exhibiting both electric and magnetic dipolar resonances (Figure 4(a)). The nanodisks were fabricated using a silicon-on-insulator (SOI) wafer, and the sharpest effect in THG enhancement was demonstrated around their magnetic dipole resonance, for which the silicon metasurface generated up to $2 \, nW$ THG power for a pump power of $30 \, mW$. The resulting conversion efficiency of $0.9 \times 10^{-7}$ was fundamentally limited by two-photon absorption (free carrier generation) in the Si substrate, which is manifested in the appearance of saturation at the peak pump intensity of $5 \, GW/cm^2$. These results, obtained in the "transmission" configuration, might be improved in the "reflection" regime, since the saturation regime can be shifted to larger pump intensities in this case.

Resonant high-index dielectric nanoparticles provide wide opportunities to design more complex systems possessing *Fano resonances* [226]–[232]. The Fano resonance originates from the interference of at least one resonant and one nonresonant scattering channels, and it manifests itself as a non-symmetrical dip in the scattering spectrum. At the Fano resonance frequency, the near field of the resonant nanostructure is enhanced, thanks to the destructive interference of the resonant and nonresonant modes, providing a powerful approach to increase light-matter interaction and nonlinearities. In addition to a strong local field enhancement, the asymmetric profile of the Fano resonance in such structures allows to control the radiative damping of the resonant modes. Interestingly, that all-dielectric nanostructures exhibit not only an electric type of Fano resonance, but also a magnetic one, which is associated with the optically induced magnetic dipole mode of individual high-index nanoparticle. Figure 4(b) shows a Fano nonlinear metasurface [224] consisting of resonant Si nanodisks that provide highly resonant dark modes, and Si nanoslits supporting the bright mode. The measured conversion efficiency of $1.2 \times 10^{-6}$ with average pump power of $50\,\text{mW}$ at peak pump intensity of $3.2\,\text{GW}/\text{cm}^2$ shows the improvement associated with a Fano response in the metasurface. The enhanced nonlinearity, combined with a sharp linear transmittance spectrum, results in transmission modulation with a modulation depth of $36\%$, which was demonstrated with a pump-probe experiment. The Fano-assisted third-harmonic generation in trimers and quadrimers were also demonstrated in Ref. [233] and Ref. [234], respectively. Extensive theoretical studies on third-order nonlinear properties of high-index dielectric nanostructures have been presented in Ref. [235].

As mentioned in the introduction, one of the main advantages of high-index all-dielectric structures is their high damage threshold, which enables operation with extremely large pump intensities and femtosecond laser excitation. The subwavelength thicknesses of metasurfaces allows generating femtosecond pulses at second and third harmonic without additional pulse compression. This application of all-dielectric metasurfaces was explored in Ref. [5]. In this work, a novel approach to produce these nonlinear metasurfaces based on self-organization of Si thin film upon intense femtosecond laser radiation was proposed. The resulting self-assembled metasurface is tuned to operate at a wavelength of choice, as a function of the laser employed in the fabrication. The proposed approach has been used to fabricate a nonlinear metasurface with 30-fold enhanced third-order nonlinear response, demonstrating the generation of UV femtosecond laser pulses at a wavelength of 270 nm. Moreover, the produced Si metaatoms located on a Si thin film can provide additional thermal sink. As a result, the metasurface has sufficient damage threshold, and can withstand up to $10^3 \text{ GW}/\text{cm}^2$.

High-index dielectric nanoparticles are also of interest because of other unusual electromagnetic scattering modes that they can support. For instance, recently it has been demonstrated that so-called *anapole modes* can be excited within such nanoparticles [212]. The term anapole means "without poles" in Greek, and it is characterizes by a specific configuration of excited fields inside a system when the dipole and toroidal moments being excited cancel each other in the far zone of the system radiation [212], [220], [236]–[239]. Lack of scattering and radiation loss in a dipole channel can further enhance the local fields, boosting nonlinear effects. In Ref. [102], it has been demonstrated that THG in thin Ge nanodisks under normally incident laser excitation can be boosted via this nonradiative anapole mode, Figure 4(d). The metasurface in this case was not optically dense, thus the nonlinear response of the entire structure was mostly obtained by the properties of a single nanodisk. At the anapole wavelength, the scattering of the nanoparticle is strongly suppressed; at the same time, the confinement of the electric field inside the disk increases, giving rise to strong nonlinearities. Figure 4(d) shows the measured THG power versus pump power at the anapole resonance. The left inset shows the corresponding THG intensity taken at $\lambda_{\text{FF}} = 1650 \text{ nm}$. The observed conversion efficiency $\approx 10^{-4}$ upon $1\,\mu\text{W}$ ($15 \text{ GW}/\text{cm}^2$) pump power is about one order of magnitude larger than in the Fano metasurface discussed above (see Figure 4(b)).

The previous discussion shows that Si-based nanostructures and metasurfaces provide strong THG properties. However, due to the centrosymmetrical crystal structure of silicon, second-order nonlinear optical phenomena were not observed in Si-based metasurfaces. Since SHG requires symmetry breaking, it can typically occur only in the volume of non-centrosymmetric materials or at a material interface [92], [160]. This problem has been overcome using nanoscale resonators made out of III-V semiconductors [225]. Currently, III-V semiconductors are of particular interest in nanophotonics because they have high internal quantum efficiency and offer the possibility to form high-quality heterostructures [225], [240]. Moreover, III-V semiconductors such as GaAs and AlGaAs possess high dielectric index and relatively large second-order susceptibility $\chi^{(2)} \approx 200\,\text{pm/V}$ [94], [225]. For instance, in Ref. [225] a dielectric metasurface based on GaAs cylindrical nanoparticles demonstrating high performance for SHG has been presented. The strongest SHG effect was observed at the magnetic dipole resonance, at which the absolute nonlinear conversion efficiency reaches $\approx 2\times10^{-5}$ at $3.4\,\text{GW/cm}^2$ pump intensity, see Figure 4(c). At the same time, the demonstrated conversion efficiency at the magnetic dipole resonance is about 100 times higher than the conversion efficiency at the electric dipole resonance, which is caused by increased absorption of GaAs at the shorter wavelength of the electric dipole resonance.

Shaping the radiation and polarization patterns of SHG from all-dielectric nanostructures (single AlGaAs nanodisks), including achieving unidirectional harmonic generation and nonlinear generation of beams of complex polarization have been experimentally realized in Ref. [94]. In this paper, the nanoscale sources of SHG emitting vector beams with designed radial polarization and conversion efficiencies exceeding $10^{-4}$ have been demonstrated.

Concluding this section, we would like to emphasize that the high-index dielectric metasurfaces provide a strong nonlinear response, low dissipative losses and high damage threshold. These advantages make them a powerful platform for modern nonlinear nanophotonics. The presence of electric and magnetic responses makes it possible to tune the scattering power patterns and design switchable flat optical devices engaging these nonlinearities. Perovskite nanostructures can also possess Mie electric and magnetic resonances in the optical range, and they may provide another exciting material opportunity for dielectric nonlinear metasurfaces, given their strong nonlinear response. We discuss these opportunities in the Outlook section.

## IV. Semiconductor-Loaded Nonlinear Plasmonic Metasurfaces

In the previous sections, we have shown that plasmonic and all-dielectric metasurfaces provide relatively large conversion efficiencies stemming from concentrated fields in small volumes. Strong nonlinear generation emerges either due to the excitation of surface nonlinearities by extremely strong fields in specifically designed plasmonic particles and nanoantennas, or from large intrinsic nonlinearities pumped at very high intensities. It is desirable to combine strong field enhancement provided by plasmonic resonances with large intrinsic nonlinear susceptibility of novel materials to push the conversion efficiency to even larger values. Various types of combined structures, consisting of metallic resonators engaging strong fields in very small volumes, loaded with nonlinear dielectric media employing these strong fields for efficient nonlinear conversion, have been proposed and realized in recent years. In particular, several orders of magnitude increase in nonlinear conversion efficiencies of second- and third- order effects have been demonstrated in metal/dielectric core-shell nanoparticles [241]–[243], plasmonic SRRs placed top of nonlinear GaAs substrates [244], nanopatterned plasmonic films filled with GaAs [245], and dipole antennas loaded with nonlinear media [92], [112], [133], [241], [246]–[248]. Nineteen orders of magnitude increase in FWM efficiency has been shown in a metasurface consisting of nm-thin resonant gaps formed between silver nanorods and a thick substrate, filled with a $\chi^{(3)}$ nonlinear medium [198]. In this case, the bulk nonlinearity stemming from naturally available nonlinear media has been employed. Importantly, the advantage of employing plasmonic metasurfaces is not only limited to field enhancement, but it extends into their ability to tailor and control the field polarization and phase at the nanoscale [155]. As we show below, this property is of invaluable importance to exploit the unique nonlinear properties of metasurfaces which are able to confine strong nonlinear responses in subwavelength volumes, and they can be at the same time employed to control the phase of the nonlinear signal with the high spatial resolution and precision, impossible when considering phase matching constraints in conventional nonlinear optical devices.

As was briefly discussed in the introduction, MQWs are artificially designed multilayer media consisting of III-V semiconductor stacks. This class of artificial semiconductor materials allows a wide flexibility in engineering the bandgaps, and it is routinely used to form heterojunctions – boundaries between different semiconductor media with abrupt changes in the binding energy. By combining several heterojunctions, we can construct a set of potential wells for charge carriers in the transverse direction. Due to this confinement, the energy of electrons and holes moving orthogonally to the layers becomes quantized. A single quantum well formed by two GaAs/AlGaAs heterojunctions and the corresponding electron-hole band structure are schematically shown in Figure 5(a). If the width of the potential well formed by the material with the smaller band gap is less than the de Broglie wavelength of the electrons (holes) in the material (about 15 nm for electrons in GaAs), the carrier motions will be quantized in the direction normal to the growth axis ($z$), whereas in the plane of the quantum wells ($x$-$y$), their motion remains free. By growing more semiconductor layers, adjusting their widths and doping levels, that can be done with high precision using modern semiconductor growth techniques, we can precisely engineer interband and intersubband transition rates and frequencies between them, effectively quantum-engineering the frequency response of the material. In particular, by tuning the energy levels in GaAs/AlGaAs heterostructures and combining multiple quantum wells, see Figure 5(b), one can engineer giant levels of effective $\chi^{(2)}$ and $\chi^{(3)}$, many orders of magnitude larger than in conventional nonlinear media [41], [43]. Due to the fact that electron-hole motion along the layers is free, the nonlinearity in MQWs can be engaged only by an electric field applied orthogonally to the layers (here, $z$-direction), and all transverse nonlinear susceptibility tensor elements are zero. Direct excitation of MQWs by light impinging orthogonally to the semiconductor layers is effectively not possible. MQWs also possess a large index $n_{MQW} \sim 3$, with significant anisotropy and absorption around the frequencies close to the intersubband transitions responsible for the large nonlinear response, making efficient coupling and outcoupling of light with the right polarization even more challenging.

The possibility of inducing large second-harmonic generation in MQWs was explored originally in AlGaAs heterostructures with asymmetric composition gradients, which would provide the required symmetry breaking condition [250]. A voltage bias may be used to modify and spectrally tune intersubband nonlinearities [44], [251], and electrical pumping may be used to produce active intersubband structures with full loss-compensation for second- [252], [253] and third-order [254] nonlinear processes. However, due to the fact that induced nonlinear polarization currents in MQWs are also oriented orthogonally to the layers, all these structures can radiate only side-wise. In such a geometry, a significant fraction of generated photons is lost due to the absence of longitudinal phase-matching, and only a part of the volume near the outcoupling aperture is actually able to radiate efficiently.

In order to facilitate the coupling of normally impinging light and convert it into $z$-polarized electric field within the MQW volume, the use of plasmonic metasurfaces consisting of dogbone- or SRR-shaped gold nanoresonators placed on top of MQWs has been proposed [207], [255], [256]. These experimental studies confirmed that plasmonic metasurfaces indeed enable efficient light coupling to intersubband transitions in MQWs over a very wide range of incidence angles. Almost at the same time, it has been shown that double resonant structures, such as asymmetrically crossed dipoles and SRRs, can also be used to enable efficient radiation from the *z*-oriented second-harmonic polarization currents, ultimately leading to 400nm-thick MQW-based metasurfaces (shown in Figure 5(c)) with SHG efficiency $\eta_{SHG} \sim 10^{-6}$ at pump intensities of only $15\,\text{kW}/\text{cm}^2$ [183], [257], 8 orders of magnitude larger than in any previously fabricated structure. The record-high SHG conversion efficiency of 0.075 % has been experimentally achieved at similar pump intensities using T-shaped gold resonators with MQWs etched around, as depicted in Figure 5(d) [249]. The structure has been optimized to provide the largest overlap integral (2) between *z*-polarized fields at $\omega$ and $2\omega$ induced in MQW (see Figure 5(e)) within a small footprint. MQW etching enables each unit-cell to operate as a plasmon-dielectric nanocavity, further promoting conversion among *x*- and *y*-polarized pump and SH radiation, respectively, and *z*-polarized fields inside MQW volumes.

We envision that even larger SHG conversion efficiencies, on the order of a few percent might be achieved with suitably designed and higher quality MQWs, and proper optimization of the plasmonic structures. The main constraint in these metasurfaces comes from the relatively fast saturation of the nonlinear response as the input intensity grows, due to very strong local fields in MQW volumes. As seen in Figure 5(f), even at low pump intensities the SH power $P_{SHG}$ is not linearly proportional to the square pump power $P_{FF}$, as it would have been in the absence of saturation. A detailed theoretical analysis of MQW-loaded plasmonic metasurfaces revealed that for this class of structures there is a critical pump intensity beyond which saturation effects start to dominate, and eventually suppress the nonlinear response [184].

## V. Phase Control over Nonlinear Metasurfaces

The strong interest in metasurfaces and metasurface-based flat optical devices is explained by the fact that they enable a large degree of control over the scattered fields by introducing abrupt discontinuities in the impedance, induced by specifically tailored transverse impedance gradients. Reverse engineering of the required impedance profile, i.e., calculating the necessary currents that may sustain the desired spatial field profile, is directly applicable to these structures [258]. In linear optics, proper engineering of subwavelength meta-atoms enabled a wide variety of functionalities, such as focusing/defocusing, steering in arbitrary directions, as well engineering advanced phase front patterns needed to produce helical beams carrying non-zero orbital angular momentum (OAM) or computer-generated holograms [155], [259]–[262]. In this Section, we discuss how similar concepts can be extended to nonlinear metasurfaces, in order to achieve full control over high-harmonic fields. This is unique to the nonlinear metasurface paradigm, because in conventional nonlinear optics approaches the large volume inherently required to sustain a non-negligible nonlinear response hinders the achievable spatial resolution in controlling the nonlinear fields. In turn, nonlinear metasurfaces provide a unique platform to boost the effects and confine them within deeply subwavelength unit cells that can be independently tailored.

In the linear metasurface paradigm, the precise control over the local scattered field phase enables the tailoring of the far-field radiation pattern. A large number of strategies to gain control over the phase distribution have been devised in recent years, such as changing the size, orientation, aspect ratio, shape, and arrangement of elements [155]. In the nonlinear regime, generated fields turn out to be very sensitive to variations in the phase and amplitude of the fields at pump frequency, and small errors may lead to large phase discrepancies and amplitude jumps. The fact that even at normal incidence nonlinear metasurfaces often can only radiate towards oblique angles further complicates the translation of linear metasurface approaches to the nonlinear regime.

It has been recently shown that some phase control functionality can be achieved in thin layers of conventional nonlinear media applying special poling patterns [263]. Poling techniques, consisting in reversing the sign of $\chi^{(2)}$ or $\chi^{(3)}$ in a specific spatial pattern in order to imprint a momentum along specific directions, is routinely used to achieve quasi-phase-matching in bulk nonlinear crystals [1]. In Ref. [263], poling is applied orthogonally to the pump beam propagation direction, so that the additional momentum pattern is transferred to the second-harmonic signal to realize different patterns, including circularly polarized (CP) vortex beams (diffraction orders) carrying different OAMs and propagating in distinct directions.

Another approach to gain control over the phase of the nonlinear generated field has been demonstrated in nanopatterned Au films [120]. By changing the aspect-ratio of the nanoholes, the phase of the FWM signal radiated in forward direction can be tailored with subwavelength precision while maintaining the same FWM field amplitude. Using this approach, beam-steering and focusing functionalities have been experimentally realized. However, this technique is not applicable to second-order processes, for which local symmetry breaking is required. A different approach, conceptually similar to the aforementioned transverse poling and consisting in a periodic change of the SHG signal sign by flipping the SSR orientations has been used to realize vortex-beam shaping in metallic nonlinear metasurfaces [121].

Recently, it has also been shown that, instead of using simple binary phase patterns, a more flexible and efficient technique to arbitrarily tailor the nonlinear phase with high resolution is available in the form of Pancharatnam-Berry optical elements [118]. These are polarization-sensitive elements, such as SRRs, whose orientation is gradually varied along the metasurface, as shown in Figure 6(a,c). Under circularly polarized pumping light, the local orientation gradients add different phases to the two CP states of the nonlinear field. Because these phases depend solely on the elements' orientation, it can be interpreted as being of geometrical nature and it is often referred to as geometrical phase [264], [265]. This approach has been often used to control the scattered field in linear metasurfaces [180], [266]–[269]. The geometrical phase approach has been successfully applied to MQW-loaded nonlinear metasurfaces aimed at SHG and THG (similar to the ones discussed in Section IV) to achieve beam-steering, focusing/defocusing and vortex-beam shaping functionalities [118], [119], [270], as shown in Figure 6(c,d). It is important to mention that such metasurfaces provide both, large efficiency *and* subwavelength phase control, simultaneously, without the need of sacrificing the performance.

In a general theoretical framework employing the effective nonlinear susceptibility, as in (2), it has been shown that for sufficiently smooth spatial variation of the SSR orientation, $\varphi(x, y)$, the local susceptibility tensor in a circular polarization basis $l, m, n = \{R, L, z\}$ (RCP, LCP, and $z$-polarization, respectively) can be found as [185]

$$\chi^{(2)}_{\text{eff},lmn}(x, y) = \chi^{(2)}_{\text{eff},lmn} e^{i(l-m-n)\varphi(x,y)}, \qquad (4)$$

where $\chi^{(2)}_{\text{eff},lmn}$ is evaluated for a reference orientation, and $l$, $m$, and $n$ in the exponent correspond to $-1$ for right-handed polarization ($R$), to $+1$ for left-handed polarization ($L$), and 0 for $z$-polarization. From Eq. (4) it follows, for example, that $\chi^{(2)}_{\text{eff},RRR}(x, y) = \chi^{(2)}_{\text{eff},RRR} e^{i\varphi(x,y)}$, $\chi^{(2)}_{\text{eff},RLL}(x, y) = \chi^{(2)}_{\text{eff},RLL} e^{-i3\varphi(x,y)}$, $\chi^{(2)}_{\text{eff},RLz}(x, y) = \chi^{(2)}_{\text{eff},RLz} e^{-i2\varphi(x,y)}$, and so on. It is clear that in the CP basis every combination of polarizations of the impinging and outgoing beam *lmn* acquires a specific phase factor that depends linearly on the local orientation $\varphi(x, y)$ of each element, with a 360 deg phase range readily available. By tailoring $\varphi(x, y)$, i.e., rotating each element in the array by a certain angle, one can simultaneously manipulate the CP second-harmonic radiation components with subwavelength resolution. A similar approach can also be applied to other nonlinear processes, such as THG [185]. MQW-based nonlinear metasurfaces, with their unparalleled level of conversion efficiencies per unit volume, and equipped with advanced field control functionalities, have been opening a new paradigm for nonlinear optics, requiring much simpler optical setups to arbitrarily control with subwavelength resolution the nonlinear fields and opening an exciting path to compact and robust nonlinear conversion and other nonlinear processes.

## VI.     Summary and Outlook

In this paper, we have reviewed the state-of-the-art in the intensely developing area of nonlinear metasurfaces, including those based on plasmonic and high-index dielectric nano-inclusions, as well as semiconductor-loaded plasmonic metasurfaces with special emphasis on multi-quantum-well-based nanostructures. We have discussed the role of optical modes in harmonic generation, including magnetic and anapole moments. We have also considered the phase control of generated harmonics by engineering the transverse impedance gradients along the nonlinear metasurfaces, focusing on the Pancharatnam-Berry approach. To summarize all results presented in the Review, we have put together in Table 1 a summary of various approaches to the creation of nonlinear metasurfaces for SHG and THG, in order to facilitate a direct comparison of the proposed methods. For comparison, the parameters of $LiNbO_3$ are also shown.

We note that plasmonic nanostructures provide unique opportunities for light localization at the nanoscale, simultaneously becoming a powerful tool to multipole mode composition engineering. All these advantages make plasmonic metasurfaces an interesting platform for nonlinear nanophotonics. However, these metasurfaces possess low damage threshold, which is caused by dissipative loses of plasmonic nanoinclusions. In contrast, high-index dielectric nanostructures support high damage threshold, providing the opportunity of using stronger pump signals. This makes all-dielectric metasurfaces very appealing for high-intensity applications such as generation of femtosecond pulses. III-V semiconductors provide strong second-order nonlinear response, expanding the use of high-index nanoparticles to second-order nonlinear effects.

Table 1: Comparative table of nonlinear metasurface performance based on different approaches

| Nonlinear medium | max $\eta_{SHG}$ | $\chi^{(2)eff}$, m/V | max $\eta_{THG}$ | $\chi^{(3)eff}$, m$^2$/V$^2$ | Phase control | Advantages | Disadvantages | Ref. |
|---|---|---|---|---|---|---|---|---|
| LiNbO$_3$ | $10^{-10}$…$10^{-11}$ | ~$10^{-12}$ | – | ~$10^{-18}$ | No | Commercially available | Relatively weak nonlinearity | [271] |
| Plasmonic metasurfaces | $10^{-9}$…$10^{-10}$ | $10^{-18}$ | ~$10^{-2}$ | $10^{-19}$ | Yes | Strong field enhancement | Low optical damage threshold | [122], [272] |
| Dielectric metasurfaces | $10^{-6}$…$10^{-7}$ | ~$10^{-12}$ | ~$10^{-6}$ | ~$10^{-15}$ | Yes | High optical damage threshold | Moderate field enhancement | [219], [220], [234], [273], [274] |
| MQW-based metasurfaces | $10^{-4}$ | $10^{-6}$ | - | - | Yes | Extremely high efficiency for low input intensities | Low saturation intensity | [119], [183], [249] |

We would like to emphasize that the metasurfaces based on MQW structures coupled with plasmonic nanostructures have recently demonstrated record values of $\eta_{SHG}$. The nonlinear optical response of these artificial surfaces can be specifically designed for operation at any wavelength of the pump signal over a wide spectral range. However, MQW-based metasurfaces possess a relatively low saturation threshold, which hinders the application of these structures in optical systems with large pump intensities. The future quest for giant optical nonlinearity in metasurfaces may be based on the search for novel material substrates.

One of such family of materials that has been attracting more and more attention is the perovskites, such as BaTiO$_3$, CsPbBr$_3$, CsBi$_{1-x}$Eu$_x$Nb$_2$O$_7$ ($x = 0, 0.1,$ and $0.2$). These materials are used for modern color converters for visible light communication, photovoltaic and light-emitting devices, sensors, high-performance permittivity materials, nonlinear nanophotonics and high-harmonic generation. Recently, it has been demonstrated that individual barium titanate (BaTiO$_3$) nanoparticles can also possess magnetic Mie resonances [27]. Moreover, SHG enhancement in the range of four orders of magnitude within the same nanoparticle has been demonstrated in the visible range. In contrast to plasmonic and semiconductor nanostructures without centrosymmetry, SHG in $BaTiO_3$ nanoparticles originates from the bulk, therefore the enhancement of electric field inside the nanoparticle can further increase the conversion efficiency. It is important to point out that $BaTiO_3$ nanoparticles are bio-compatible, which makes them particularly appealing for biophotonics and nonlinear bioimaging applications.

## Acknowledgements

The authors are thankful to Dr. Denis Baranov for useful discussions. The authors declare no competing financial interest. We acknowledge the support of the Air Force Office of Scientific Research, the Welch Foundation with grant No. F-1802, and of the National Science Foundation.

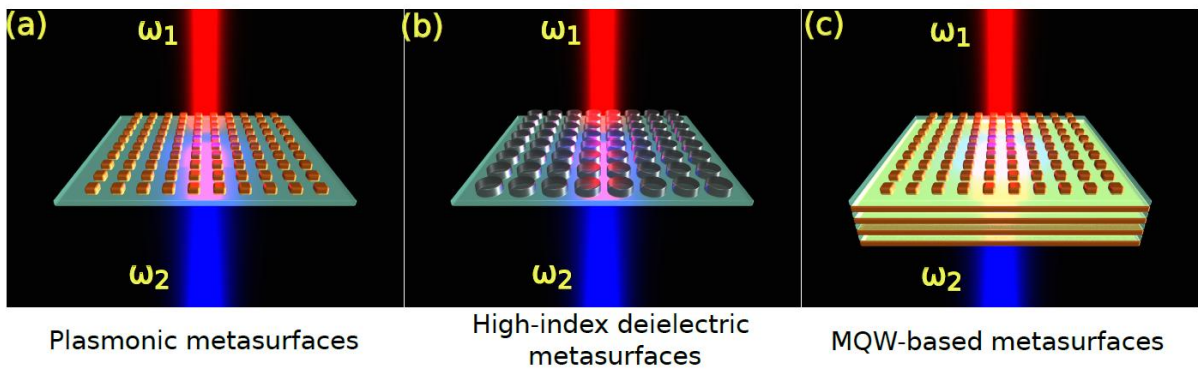

Figure 1: Schematic of thin metasurfaces for flat nonlinear nanophononics, consisting of (a) plasmonic all-metallic meta-atoms, (b) high-index all-dielectric meta-atoms, and (c) plasmonic resonators loaded with multi-quantum-wells.

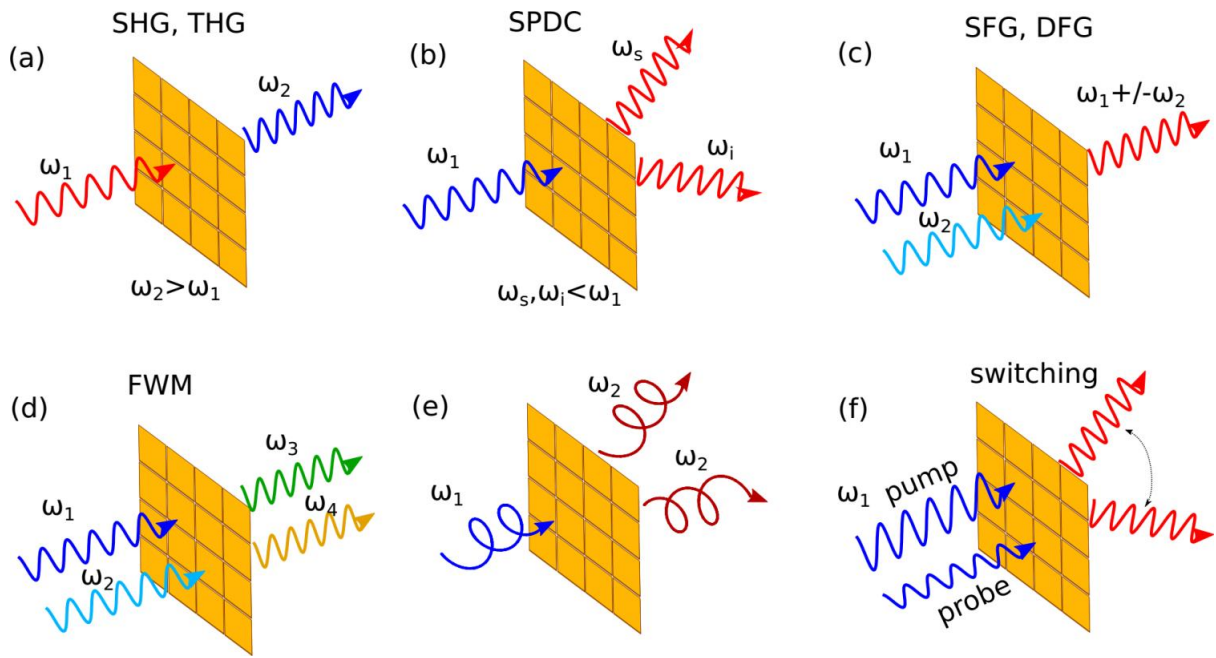

Figure 2: Selected functionalities of nonlinear metasurfaces: (a) second- (SHG) and third-harmonic generation (THG) , (b) spontaneous parametric down-conversion (SPDC), (c) sum- (SFG) and difference-frequency generation (DFG), (d) four-wave mixing (FWM), (e) nonlinear phase control, (f) nonlinear switching and

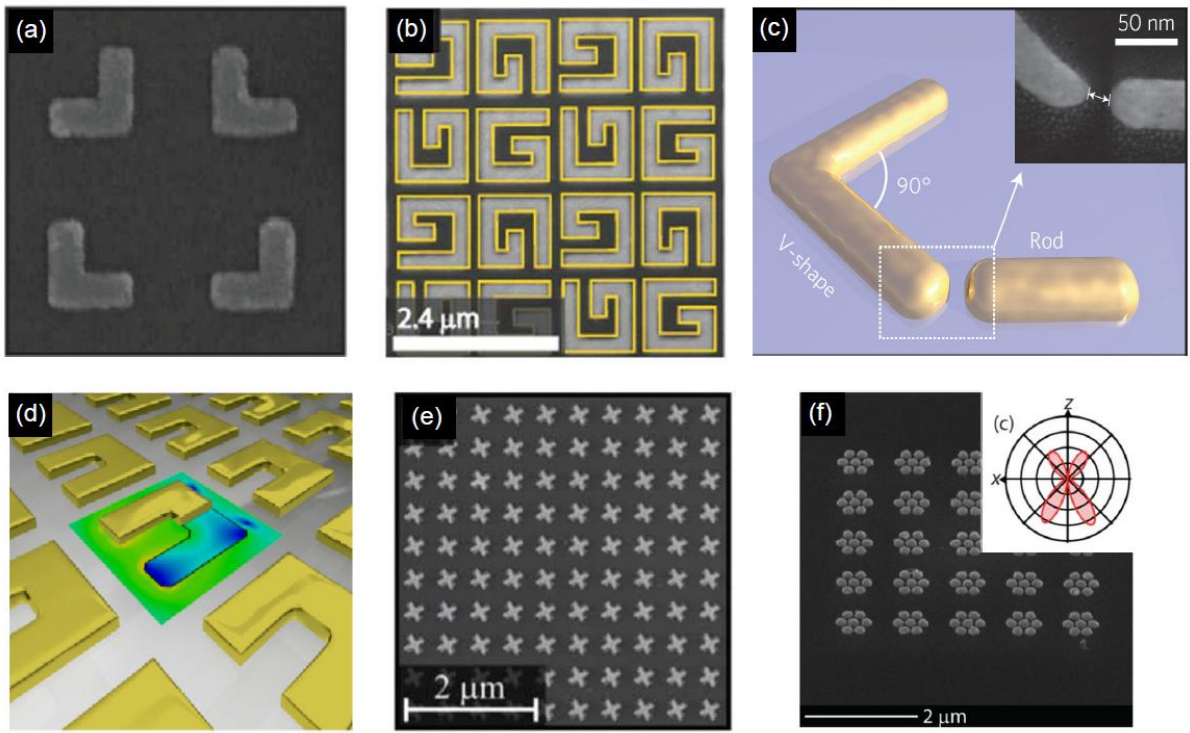

Figure 3: Plasmonic metasurfaces exhibiting strong second-harmonic generation. Breaking of centrosymmetry is achieved by using asymmetric plasmonic structures, such as (a) L-shaped nanoantennas, (b) chiral G-shaped structures, (c) multi-resonant antennas, (d) split-ring resonators. Periodic metasurfaces consisting of centrosymmetric structures can also generate second-harmonic radiation, (e) being illuminated by circularly polarized light, or (f) excited and (or) observed at oblique angles. The far-field directivity plot shown in the inset clearly shows that the structure cannot radiate normal to the metasurface due to symmetry constraints, even in the case of strong local SH fields. Figure adapted with permission from: (a) – Ref. [85]; (b) – Ref. [161] (c) – Ref. [91]; (d) – Ref. [86]; (e),(f) – Ref. [186].

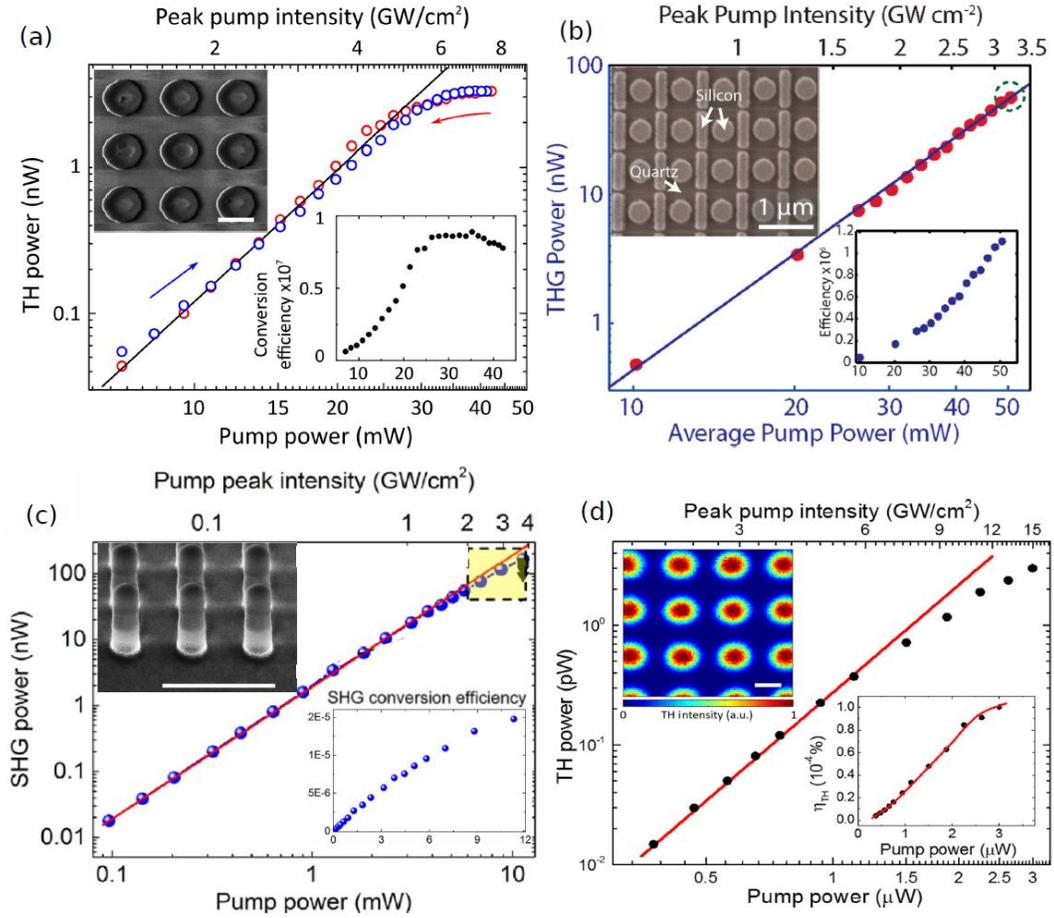

Figure 4: (a) Power dependence and conversion efficiency of resonant THG process in Si nanodisks. Blue circles denote the THG power dependence upon increasing the pump power, while red circles denote the reverse procedure obtained at $\lambda = 1260$ nm (magnetic resonance) fundamental wavelength. Right inset: conversion efficiency as a function of pump power; left inset: the SEM image of the sample (scale bar is $500$ nm). (b) Log plot of the third harmonic power as a function of pump power and peak pump intensity. The red circles indicate the measured data, and the blue line is a numerical fit to the data with a third-order power function. Right inset: extracted absolute THG efficiency as a function of pump power; left inset: SEM image of the Fano-resonant metasurface. (c) Quadratic relationship between average pump and SHG powers at low pump intensities, and the deviation from the quadratic relationship at higher pump intensities due to the damage of GaAs resonators. Left inset: a 75° side view SEM images of the fabricated GaAs dielectric resonator array. Right inset: SHG conversion efficiency as a function of pump power. (d) Measured THG power versus pump power at the anapole mode. Left inset: corresponding THG intensity image taken at $\lambda_{FF} = 1260$ nm; scale bar is $1 \mu m$. Right inset: conversion efficiency, $\eta_{THG}$, as a function of pump power. Figure adapted with permission from: (a) – Ref. [219]; (b) – Ref. [275]; (c) – Ref. [225]; (d) – Ref. [102].

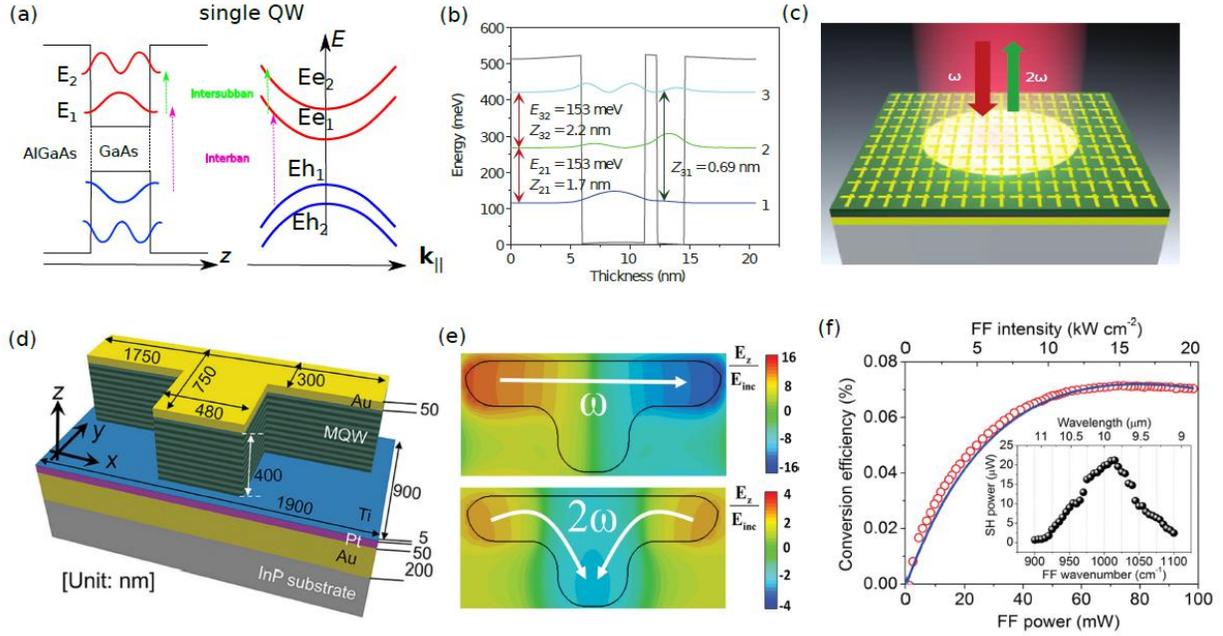

Figure 5: Second harmonic generation with plasmonic metasurfaces coupled to intersubband transitions of multi-quantum-well structure. (a) Energy versus the direction ($z$), normal to the growth axis, for a single QW based on AlGaAs/GaAs semiconductors. (b) Specifically, designed conduction band diagram for one period of an $In_{0.53}Ga_{0.47}As/Al_{0.48}In_{0.52}As$ MQW designed for giant nonlinear response for SHG. (c) Left: Schematic of the metasurface design and operation. Red and green arrows indicate the incident pump beam at fundamental frequency $\omega$ and the reflected second-harmonic beam at frequency $2\omega$, respectively; Right: Scanning electron microscope images of the fabricated metasurface. (d) A unit-cell of the nonlinear metasurface structure. (e) Top view of the normalized $E_z^\omega$ and $E_z^{2\omega}$ field distributions in the T-shaped resonator. (f) Experimentally measured (red circles) and theoretically calculated (blue solid line) SHG power conversion efficiencies as a function of the fundamental peak power (bottom axis) or peak intensity (top axis) at wavenumber of 1015 cm$^{-1}$ for *yxx* polarization combination. Inset: SH peak power as a function of pump wavenumber for *yxx* polarization combination (pump peak power is 35 mW). Figure adapted with permission from: (b), (c) – Ref. [183]; (d)-(f) – Ref. [249].

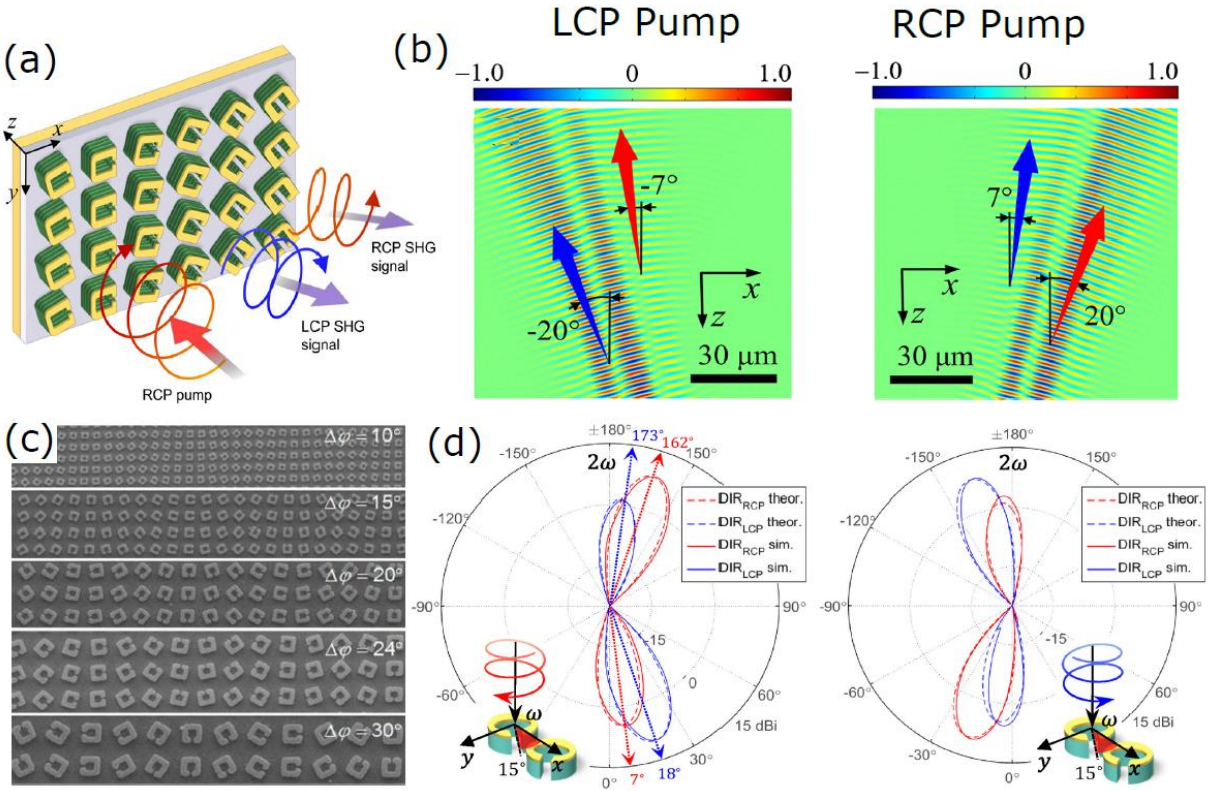

Figure 6: Second harmonic generation with gradient plasmonic metasurfaces coupled to intersubband transitions of multi-quantum well structure. (a) A general sketch of a nonlinear metasurface with a orientation gradient (in the x direction.) (b) Spatial distribution of the *y* component of the radiated field above the metasurface illuminated by a 30 μm-wide LCP (left figure) or RCP (right figure) Gaussian beam (the incident field is not shown). (d) SEMs of fabricated gradient SRR arrays, with differing angular rotational steps. (e) Far-field directivity patterns (in dBi) for RCP and LCP components of SH radiation generated by a nonlinear metasurface illuminated by a Gaussian beam of 12 μm full-width at half-maximum. The directivity plots are shown in the *x-z* plane. Theoretical results are shown with dashed lines, and solid lines show results from full-wave simulations. Figure adapted with permission from: (a),(b) – Ref. [118]; (c) – Ref. [119]; (e), (f) – Ref. [185].